# Efficient quantum secure direct communication with authentication[*]

**Liu Wen-jie(刘文杰)**[a,b**], **Chen Han-wu(陈汉武)**[a], **Li Zhi-qiang(李志强)**[a], **Liu Zhi-hao (刘志昊)**[a]

[a] School of Computer Science and Engineering, Southeast University, Nanjing 210096, China
[b] School of Computer & software, Nanjing University of Information Science & Technology, Nanjing 210044, China

Two protocols of quantum direct communication with authentication [Phys. Rev. A 73, 042305(2006)] were recently indicated to be insecure against the authenticator Trent's attacks [Phys. Rev. A 75, 026301(2007)]. We present two efficient protocols by using four Pauli operations, which are secure against inner Trent's attacks as well as outer Eve's attacks. Finally, we generalize them to multiparty quantum direction communication.

Since Wiesner [1] and then Bennett et al.[2] found that quantum effects can be used to transmit secrete information in an open quantum channel, a remarkable surge of interest in the international scientific and industrial community has propelled quantum cryptography communication into mainstream of computer science and physics. Quantum key distribution (QKD) is the earliest and most mature branch of the mainstream, which provides a novel way for two legitimate parties to share a common secret key over a long distance with negligible leakage of information to the eavesdropper. Its ultimate advantage is the unconditional security. Hence, after Bennett and Brassard published BB84 protocol [3], a variety of QKD protocols have been proposed, such as E91 [4], B92 [5], and six-state protocol [6] etc. And the security of some QKD protocols was theoretically proven [7-9].

Recently, a new concept of quantum cryptography communication, quantum secure direct communication (QSDC), was proposed. Different from QKD, the deterministic QSDC protocol is to transmit directly the secret message without generating a random key to encrypt them, so it is more demanding on security. In 2002, Beige et al. presented the pioneering quantum secure direct communication scheme with singlet photon [10], then Bostrom and Felbingeer put forward a "Ping-pong" protocol with EPR state [11]. Since then, many other QSDC protocols have been proposed [12-24]. These QSDC protocols can protect against eavesdropping attack called passive attack in cryptography [25], while they cannot resist other kind of attack, named active attack, such as the impersonation attack, the man-in-the-middle attack etc. To make up these shortcomings, recently Lee, Lim and Yang proposed two QSDC protocols of quantum direct communication with authentication (Lee-Lim-Yang protocols) [26]. However, Zhang, Liu and Wang et al. indicated that these protocols were insecure against the authenticator Trent's attacks, and put forward two modified protocols by using the operation $\sigma_z$ instead of the operation $X$ (Zhang-Liu-Wang protocols) [27].

In this letter, we improve Lee-Lim-Yang protocols and Zhang-Liu-Wang protocols, and propose two novel efficient QSDC protocols with authentication (EQSDCPs). Our revised EQSDCPs have no need to perform the Hadamard operation $H$. They get twice transition efficiency by utilizing four kinds of Pauli operation ($I$, $\sigma_x$, $i\sigma_y$, $\sigma_z$), and transmit two bits every GHZ state. What is more, these two EQSDCPs show the same security as Lee-Lim-Yang protocols when Trent is honest, and also are secure against Zhang-Liu-Wang's Trent attacks when Trent is dishonest.

Let us start by illustrating some notations. We firstly define eight GHZ states as

$$|P^{\pm}\rangle = \frac{1}{\sqrt{2}}(|000\rangle \pm |111\rangle), \quad |Q^{\pm}\rangle = \frac{1}{\sqrt{2}}(|001\rangle \pm |110\rangle),$$
$$|R^{\pm}\rangle = \frac{1}{\sqrt{2}}(|010\rangle \pm |101\rangle), \quad |S^{\pm}\rangle = \frac{1}{\sqrt{2}}(|011\rangle \pm |100\rangle), \quad (1)$$

and four Bell states (EPR pairs) as

$$|\phi^{\pm}\rangle = \frac{1}{\sqrt{2}}(|00\rangle \pm |11\rangle), \quad |\psi^{\pm}\rangle = \frac{1}{\sqrt{2}}(|01\rangle \pm |10\rangle). \quad (2)$$

Suppose three parties, Alice, Bob and Trent, share a GHZ triplet $|\varphi\rangle_{ATB} = |P^+\rangle_{ATB}$, where the subscripts A, T and B correspond to Alice, Trent and Bob, respectively. And $|\varphi\rangle_{ATB}$ can be rewritten as follows:

$$\begin{aligned}|\varphi\rangle_{ATB} &= \frac{1}{\sqrt{2}}(|000\rangle_{ATB} + |111\rangle_{ATB})\\ &= \frac{1}{\sqrt{2}}(|\phi^+\rangle_{AB}|+\rangle_T + |\phi^-\rangle_{AB}|-\rangle_T)\end{aligned} \quad (3)$$

where $|+\rangle = \frac{1}{\sqrt{2}}(|0\rangle + |1\rangle)$ and $|-\rangle = \frac{1}{\sqrt{2}}(|0\rangle - |1\rangle)$.

Our EQSDCPs are composed of two parts: one is the authentication process, and the other is the direct communication process with dense coding. There are three parties, Alice and Bob are the two legitimate users of the communication, and Trent is the third party who will be used to authenticate the two users. And Trent is assumed to be more powerful than the other two parties and supplies the GHZ states.

---
[*] Supported by the National Natural Science Foundation of China under Grant No. 60572071, the Natural Science Foundation of Jiangsu Province, China under Grand Nos. BM2006504 and BK2007104, and the College Natural Science Foundation of Jiangsu Province, China under Grand No. 06KJB520137.
[**] Email: wenjiel@163.com

The purpose of the authentication process is to let the three participants safely share GHZ states. Similar to Ref.[26], we suppose Trent is a trusted third party, and he shares a secret identity number $ID_{user}$ and a one-way hash function $h$ with the legitimate users. Here the hash function is

$$h: \{0,1\}^l \times \{0,1\}^m \to \{0,1\}^n, \quad (4)$$

where $l$, $m$ and $n$ denote the length of the identity number, the length of a counter, and the length of authentication key, respectively. Thus the user's authentication key can be expressed as $AK_{user}=h(ID_{user}, c_{user})$, where $c_{user}$ is the counter of calls on the hash function. When the user want to continue next authentication process, he can get a new authentication key by only adding the counter number, and does not change his *ID*. In the authentication process, Trent performs authentication with Alice by using Alice's authentication key $AK_A=h(ID_A, c_A)$, and with Bob by using Bob's authentication key $AK_B=h(ID_B, c_B)$. The brief procedure is as follows:

(1) Prerequisite. Alice and Bob register their secret identities and hash functions with Trent.

(2) Trent generates a GHZ triplets sequence $|\Psi\rangle = \{|\varphi_1\rangle, |\varphi_2\rangle, \cdots, |\varphi_n\rangle\}$, suppose

$$|\varphi_i\rangle = \frac{1}{\sqrt{2}}(|000\rangle_{ATB} + |111\rangle_{ATB}) \quad (i=1,2,\cdots,N). \quad (5)$$

(3) Trent makes unitary operations on $|\varphi_i\rangle$ with Alice's and Bob's authentication keys $AK_A$ and $AK_B$. For example, if the $i$th value of $AK_A$ is 0, then Trent makes an identity operation $I$ on Alice's particle (the first particle) of the $i$th GHZ state. If it is 1, a Hadamard operation $H$ is applied.

(4) Trent distributes the first particle, the third particle of every GHZ triplets to Alice and Bob, respectively, and keeps remaining particles for him.

(5) Alice and Bob make reverse operations of step (3) on their qubits with their authentication keys $AK_A$ and $AK_B$, respectively.

(6) Alice and Bob choose a subset of GHZ states for checking eavesdropper. If the error rate is higher than expected, then Alice and Bob resume the process. Otherwise, they can confirm that their counter parts are legitimate and the channel is secure, so they can continue the next phase, quantum direct communication process.

After above authentication process, the GHZ states are safely shared among the three parties, and two EQSDCPs will be put forward to transmit the secret message. We depict our EQSDCP 1 firstly.

In EQSDCP 1, the scenario is that Alice wants to send secret message to distant Bob, and there is a quantum channel between Alice and Bob. To achieve this task, Alice encodes the secret message on her particles and directly transmits them to Bob as follows:

(1) Alice selects randomly some particles of the remaining GHZ states after authentication, and performs on them randomly one of the four operations ($I$, $\sigma_x$, $i\sigma_y$, $\sigma_z$). The number of such particles is not big as long as it can provide an analysis of the error rate. Only Alice knows the positions of these sampling particles and she keeps them secret.

(2) Alice encodes the secret message with an error correction code (ECC), such as the Hamming code, the Reed-Solomon code, the CSS (Calderbank-Shor-Steane) code, or BCH (Bose-Chaudhuri-Hochquenghem) code, so that Bob can correct errors after transmission.

(3) Alice encodes the ECC-encoded message on the remanding particles after step(2) by performing the four unitary operations. According to our dense coding strategy, two bits correspond to a unitary operation: the bits 00 corresponds to the operation $I$, 01 to $\sigma_x$, 10 to $i\sigma_y$, and 11 to $\sigma_z$.

$$I = |0\rangle\langle 0| + |1\rangle\langle 1|, \quad \sigma_x = |0\rangle\langle 1| + |1\rangle\langle 0|,$$
$$i\sigma_y = |0\rangle\langle 1| - |1\rangle\langle 0|, \quad \sigma_z = |0\rangle\langle 0| - |1\rangle\langle 1|. \quad (6)$$

Then the GHZ states are transformed into different forms as follows.

(a) If the bits are 00, then

$$I_A|\varphi\rangle = \frac{1}{\sqrt{2}}(|000\rangle_{ATB} + |111\rangle_{ATB})$$
$$= \frac{1}{\sqrt{2}}\{|\phi^+\rangle_{AB}|+\rangle_T + |\phi^-\rangle_{AB}|-\rangle_T\} \quad (7)$$

(b) If the bits are 01, then

$$\sigma_{xA}|\varphi\rangle = \frac{1}{\sqrt{2}}(|100\rangle_{ATB} + |011\rangle_{ATB})$$
$$= \frac{1}{\sqrt{2}}(|\psi^+\rangle_{AB}|+\rangle_T - |\psi^-\rangle_{AB}|-\rangle_T) \quad (8)$$

(c) If the bits are 10, then

$$i\sigma_{yA}|\varphi\rangle = \frac{1}{\sqrt{2}}(|011\rangle_{ATB} - |100\rangle_{ATB})$$
$$= \frac{1}{\sqrt{2}}(|\psi^-\rangle_{AB}|+\rangle_T - |\psi^+\rangle_{AB}|-\rangle_T) \quad (9)$$

(d) If the bits are 11, then

$$\sigma_{zA}|\varphi\rangle = \frac{1}{\sqrt{2}}(|000\rangle_{ATB} - |111\rangle_{ATB})$$
$$= \frac{1}{\sqrt{2}}\{|\phi^-\rangle_{AB}|+\rangle_T + |\phi^+\rangle_{AB}|-\rangle_T\} \quad (10)$$

(4) After finishing all unitary operations, Alice sends all the qubits to Bob.

(5) Bob makes Bell measurement on pairs of particles consisting of his qubit and Alice's qubit.

(6) Trent measures her qubit in the $x$ basis $\{|+\rangle, |-\rangle\}$ and publishes the measurement outcomes.

(7) Bob can recover Alice's secret bits from Bob's and Trent's outcomes (shown in Table 1). For example, when Bob's measurement is $|\psi^+\rangle$ and Trent's publication is $|+\rangle$, Bob can infer that Alice performed a $\sigma_x$ operation and the bits she sent are 01. Note that these decoded bits consist of the sampling particles' bits and the secret message.

(8) Alice tells Bob the position of the sampling pairs and the type of unitary operations on them. Then Bob check the

sampling pairs that Alice had chosen, and he will get an estimate of error rate in the transmission.

(9) If the error rate is reasonably low, Alice and Bob can trust the process. Then Bob gets rid of the sample particles, and gets secret bits encoded with the corresponding ECC code. Otherwise, Alice and Bob abandon the transmission.

(10) Bob does error correction on the remanding bits, and finally gets the secret message from Alice.

In this message transmission process, even if Eve exists, but she at most get one particle from a GHZ state, and could not distinguish which operation Alice had performed, so she couldn't get any secret message.

Table 1
The corresponding relations of Alice's operation, Bob's measurement and Trent's publication in EQSDCP 1

| Trent's publication | Bob's measurement | Alice's operation |
|---|---|---|
| $|+\rangle_T$ | $|\phi^+\rangle_{AB}$ | $I$ (00) |
| $|+\rangle_T$ | $|\psi^+\rangle_{AB}$ | $\sigma_x$ (01) |
| $|+\rangle_T$ | $|\psi^-\rangle_{AB}$ | $i\sigma_y$ (10) |
| $|+\rangle_T$ | $|\phi^-\rangle_{AB}$ | $\sigma_z$ (11) |
| $|-\rangle_T$ | $|\phi^-\rangle_{AB}$ | $I$ (00) |
| $|-\rangle_T$ | $|\psi^-\rangle_{AB}$ | $\sigma_x$ (01) |
| $|-\rangle_T$ | $|\psi^+\rangle_{AB}$ | $i\sigma_y$ (10) |
| $|-\rangle_T$ | $|\phi^+\rangle_{AB}$ | $\sigma_z$ (11) |

In EQSDCP 2, the scenario is that Alice wants to send secret message to distant Bob, but there is not any quantum channel between Alice and Bob. To achieve this task, Alice encodes the message on her particles and transmits them to Trent, instead of Bob. So Trent is acting as the message transfer center.

The detailed procedure is similar to EQSDCP 1, except that Alice transmits her particles to Trent in step (4), Bob makes $x$ basis measurement in step (5), and Trent performs Bell measurement in step (6). Finally, Bob can infer the secret message according to Trent's measurement and his measurement (see Table 2).

Table 2
The corresponding relations of Alice's operation, Bob's measurement and Trent's announcement in EQSDCP 2

| Trent's announcement | Bob's measurement | Alice's operation |
|---|---|---|
| $|\phi^+\rangle_{AT}$ | $|+\rangle_B$ | $I$ (00) |
| $|\phi^+\rangle_{AT}$ | $|-\rangle_B$ | $\sigma_z$ (11) |
| $|\psi^+\rangle_{AT}$ | $|+\rangle_B$ | $\sigma_x$ (01) |
| $|\psi^+\rangle_{AT}$ | $|-\rangle_B$ | $i\sigma_y$ (10) |
| $|\phi^-\rangle_{AT}$ | $|+\rangle_B$ | $\sigma_z$ (11) |
| $|\phi^-\rangle_{AT}$ | $|-\rangle_B$ | $I$ (00) |
| $|\psi^-\rangle_{AT}$ | $|+\rangle_B$ | $i\sigma_y$ (10) |
| $|\psi^-\rangle_{AT}$ | $|-\rangle_B$ | $\sigma_x$ (01) |

As we know, Lee-Lim-Yang protocols are secure against Eve's attacks when Trent is honest, and the security proof was given in Ref [26]. Our EQSDCPs are similar to the Lee-Lim-Yang protocols when Trent is honest. So, What we need do is to prove that the EQSDCPs are secure against Zhang-Liu-Wang's Trent attacks when Trent is dishonest.

Let's start with the first Zhang-Liu-Wang's Trent attack to the first protocol, which is an intercept-resent attack. The procedure can be depicted as follows. At first, Trent intercepts the qubits that Alice attempts to transmit to Bob, and performs the $H$ operation on each qubit. Then Trent measures Alice's and his qubit in the $z$ basis $\{|0>, |1>\}$ separately. If the two outcomes are the same, then Trent can deduce that Alice had performed an $H$ operation corresponding to the bit 0. Otherwise, Alice had performed an $HX$ operation corresponding to the bit 1. So, Trent will get Alice's whole bits string including both the random bits string and the secret message. Finally Trent removes the random bits according to Alice's publication in step(8), and completely know the secret message.

According to this attack strategy, Trent intercepts Alice's particle, and respectively performs his qubit and Alice's qubit in the z basis. From these measurement outcomes, Trent can not know which unitary operations had been performed on Alice's qubits. For example, if the outcomes of Trent's and Alice's are the same, the possible operation is $I_A$ or $\sigma_{zA}$ (see equation (7), (10)). If the outcomes are different, the possible operation is $\sigma_{xA}$ or $i\sigma_{yA}$ (see equation (8), (9)).

Even if Trent performs the H operation before singlet measurement, and the states are transformed into

$$H_A I_A |\varphi\rangle = \frac{1}{\sqrt{2}} H_A (|000\rangle_{ATB} + |111\rangle_{ATB})$$
$$= \frac{1}{2}\{|000\rangle_{ATB} + |100\rangle_{ATB} + |011\rangle_{ATB} - |111\rangle_{ATB}\}$$, (11)

$$H_A \sigma_{xA} |\varphi\rangle = \frac{1}{\sqrt{2}} H_A (|100\rangle_{ATB} + |011\rangle_{ATB})$$
$$= \frac{1}{2}\{|000\rangle_{ATB} - |100\rangle_{ATB} + |011\rangle_{ATB} + |111\rangle_{ATB}\}$$, (12)

$$H_A i\sigma_{yA} |\psi\rangle = \frac{1}{\sqrt{2}} H_A (|011\rangle_{ATB} - |100\rangle_{ATB})$$
$$= \frac{1}{2}\{|011\rangle_{ATB} + |111\rangle_{ATB} - |000\rangle_{ATB} + |100\rangle_{ATB}\},$$ (13)

$$H_A \sigma_{zA} |\psi\rangle = \frac{1}{\sqrt{2}} H_A (|000\rangle_{ATB} - |111\rangle_{ATB})$$
$$= \frac{1}{2}\{|000\rangle_{ATB} + |100\rangle_{ATB} - |011\rangle_{ATB} + |111\rangle_{ATB}\}$$. (14)

As shown above, Trent is still unable to differentiate which operation had been performed by comparing the outcomes of Alice's and Trent's measurement. This means Trent cannot infer Alice's secret message, so our revised EQSDCP 1 is secure against Zhang-Liu-Wang's Trent attack.

The second Zhang-Liu-Wang's Trent attack to the protocol 2 can be depicted as follows. Trent performs an $H$ operation on each qubit he received from Alice, and then respectively measures his qubit and Alice's qubit in the $z$ basis to extract the secret message. According to the equation (7-10) and the equation (11-13), Trent can not get the secrete message whether the $H$ operation is performed or not. In short, our second EQSDCP is secure against Zhang-Liu-Wang's Trent attack too.

In addition, our protocols can be generalized to multiparty quantum direction communication. A simple example is the three-party efficient QSDC protocols (3EQSDCPs) by using four particles GHZ state. Suppose the four-particle GHZ is

$$|\varphi\rangle_4 = \frac{1}{\sqrt{2}}(|0000\rangle_{ABTC} + |1111\rangle_{ABTC})$$
$$= \frac{1}{\sqrt{2}}(|P^+\rangle_{ABC}|+\rangle_T + |P^-\rangle_{ABC}|-\rangle_T) \quad (15)$$

And the scenario is that two spatially separated parties Alice and Bob want to send messages to distant Charlie. At first, Trent distributes three particles of four-particle GHZ state to Alice, Bob and Charlie separately by utilizing the authentication process. Similarly, we can design two 3EQSDCPs. In the first 3EQSDCP, Alice encodes her message with $I_A$ (00), $\sigma_{xA}$ (01), $i\sigma_{yA}$ (10), $\sigma_{zA}$ (11), Bob encodes his message with $I_B$(0), $\sigma_{xB}$ (1), then Alice and Bob simultaneously send their particles to Charlie. After Charlie's GHZ measurement and Trent's $x$ measurement, Charlie will get the secret messages from Alice and Bob according to Table 3.

In the second 3EQSDCP, Alice and Bob simultaneously transmit their particles to Trent respectively after message encoding, and Charlie can extract the secret messages from Alice and Bob, respectively (see Table 4).

In this letter, we present two revised efficient protocols, which are more efficient than Ref. [25, 26] and secure against Zhang-Liu-Wang's Trent attacks. Finally, we generalize them to multiparty quantum secure direction communication.

Table 3
The relations of Alice's operation, Bob's operation, Trent's publication, and Charlie's measurement in the first 3EQSDCP.

| Trent's publication | Charlie's measurement | Alice's operation | Bob's operation |
|---|---|---|---|
| $|+\rangle_T$ | $|P^+\rangle_{ABC}$ | $I_A$ (00) | $I_B$ (0) |
| $|+\rangle_T$ | $|S^+\rangle_{ABC}$ | $\sigma_{xA}$ (01) | $I_B$ (0) |
| $|+\rangle_T$ | $|S^-\rangle_{ABC}$ | $i\sigma_{yA}$ (10) | $I_B$ (0) |
| $|+\rangle_T$ | $|P^-\rangle_{ABC}$ | $\sigma_{zA}$ (11) | $I_B$ (0) |
| $|-\rangle_T$ | $|P^-\rangle_{ABC}$ | $I_A$ (00) | $I_B$ (0) |
| $|-\rangle_T$ | $|S^-\rangle_{ABC}$ | $\sigma_{xA}$ (01) | $I_B$ (0) |
| $|-\rangle_T$ | $|S^+\rangle_{ABC}$ | $i\sigma_{yA}$ (10) | $I_B$ (0) |
| $|-\rangle_T$ | $|P^+\rangle_{ABC}$ | $\sigma_{zA}$ (11) | $I_B$ (0) |
| $|+\rangle_T$ | $|R^+\rangle_{ABC}$ | $I_A$ (00) | $\sigma_{xB}$ (1) |
| $|+\rangle_T$ | $|Q^+\rangle_{ABC}$ | $\sigma_{xA}$ (01) | $\sigma_{xB}$ (1) |
| $|+\rangle_T$ | $|Q^-\rangle_{ABC}$ | $i\sigma_{yA}$ (10) | $\sigma_{xB}$ (1) |
| $|+\rangle_T$ | $|R^-\rangle_{ABC}$ | $\sigma_{zA}$ (11) | $\sigma_{xB}$ (1) |
| $|-\rangle_T$ | $|R^-\rangle_{ABC}$ | $I_A$ (00) | $\sigma_{xB}$ (1) |
| $|-\rangle_T$ | $|Q^-\rangle_{ABC}$ | $\sigma_{xA}$ (01) | $\sigma_{xB}$ (1) |
| $|-\rangle_T$ | $|Q^+\rangle_{ABC}$ | $i\sigma_{yA}$ (10) | $\sigma_{xB}$ (1) |
| $|-\rangle_T$ | $|R^+\rangle_{ABC}$ | $\sigma_{zA}$ (11) | $\sigma_{xB}$ (1) |

Table 4
The relations of Alice's operation, Bob's operation, Trent's announcement, and Charlie's measurement in the second 3EQSDCP.

| Trent's announcement | Charlie's measurement | Alice's operation | Bob's operation |
|---|---|---|---|
| $|P^+\rangle_{ABT}$ | $|+\rangle_C$ | $I_A$ (00) | $I_B$ (0) |
| $|P^+\rangle_{ABT}$ | $|-\rangle_C$ | $\sigma_{zA}$ (11) | $I_B$ (0) |
| $|S^+\rangle_{ABT}$ | $|+\rangle_C$ | $\sigma_{xA}$ (01) | $I_B$ (0) |
| $|S^+\rangle_{ABT}$ | $|-\rangle_C$ | $i\sigma_{yA}$ (10) | $I_B$ (0) |
| $|P^-\rangle_{ABT}$ | $|+\rangle_C$ | $\sigma_{zA}$ (11) | $I_B$ (0) |
| $|P^-\rangle_{ABT}$ | $|-\rangle_C$ | $I_A$ (00) | $I_B$ (0) |
| $|S^-\rangle_{ABT}$ | $|+\rangle_C$ | $i\sigma_{yA}$ (10) | $I_B$ (0) |
| $|S^-\rangle_{ABT}$ | $|-\rangle_C$ | $\sigma_{xA}$ (01) | $I_B$ (0) |
| $|R^+\rangle_{ABT}$ | $|+\rangle_C$ | $I_A$ (00) | $\sigma_{xB}$ (1) |
| $|R^+\rangle_{ABT}$ | $|-\rangle_C$ | $\sigma_{zA}$ (11) | $\sigma_{xB}$ (1) |
| $|Q^+\rangle_{ABT}$ | $|+\rangle_C$ | $\sigma_{xA}$ (01) | $\sigma_{xB}$ (1) |
| $|Q^+\rangle_{ABT}$ | $|-\rangle_C$ | $i\sigma_{yA}$ (10) | $\sigma_{xB}$ (1) |
| $|Q^-\rangle_{ABT}$ | $|+\rangle_C$ | $i\sigma_{yA}$ (10) | $\sigma_{xB}$ (1) |
| $|Q^-\rangle_{ABT}$ | $|-\rangle_C$ | $\sigma_{xA}$ (01) | $\sigma_{xB}$ (1) |
| $|R^-\rangle_{ABT}$ | $|+\rangle_C$ | $\sigma_{zA}$ (11) | $\sigma_{xB}$ (1) |
| $|R^-\rangle_{ABT}$ | $|-\rangle_C$ | $I_A$ (00) | $\sigma_{xB}$ (1) |